\documentclass[
    ,final            
  ]
  {aipproc}

\layoutstyle{8x11double}

\begin{document}

\title{A Hot Spot Model for Black Hole QPOs}

\author{Jeremy Schnittman}{
  address={MIT Department of Physics, 77 Massachusetts Avenue,
Cambridge, MA 02139}
}

\author{Edmund Bertschinger}{
  address={MIT Department of Physics, 77 Massachusetts Avenue,
Cambridge, MA 02139}
}

\begin{abstract}
In at least two black hole binary systems, the Rossi X-Ray Timing
Explorer has detected high frequency quasi-periodic oscillations
(HFQPOs) with a 2:3 frequency commensurability. We propose a simple
hot spot model to explain the positions, amplitudes, and widths of the
HFQPO peaks. Using the exact geodesic equations for the Kerr metric, we
calculate the trajectories of massive test particles, which are
treated as isotropic, monochromatic emitters in their rest frames. By
varying the hot spot parameters, we are able to explain the different
features observed in ``Type A'' and ``Type B'' QPOs from XTE
J1550-564. In the context of this model, the observed power spectra
allow us to infer values for the black hole mass and angular momentum,
and also constrain the parameters of the model.
\end{abstract}

\maketitle


\section{Introduction}

One of the most exciting results from the Rossi X-Ray Timing Explorer
was the discovery of high frequency quasi-periodic oscillations from
neutron star and black hole binaries
\cite{stroh96,stroh01,lamb02}. For black hole systems, these 
HFQPOs are observed repeatedly at more or less constant frequencies,
and in a few cases with integer ratios \cite{remil02}.  These measurements 
give the exciting prospect of determining a black hole's mass and
spin, as well as testing general relativity in the strong-field
regime. To quantitatively interpret these observations in the context
of the space-time behavior near the black hole inner-most stable
circular orbit (ISCO), we have developed a ray-tracing code to model
the X-ray light curve from a collection of ``hot spots,'' small
regions of excess emission moving on geodesic orbits. The methods and
basic results of this code are described in detail in \cite{schni03}.

This hot spot model is motivated by the similarity between the
QPO frequencies and the black hole coordinate
frequencies near the ISCO \cite{stell98,stell99} as well as
the suggestion of a resonance leading to integer
commensurabilities between these coordinate frequencies
\cite{kluzn01,kluzn02}. Stella and Vietri \cite{stell99} investigated
primarily the
QPO frequency pairs found in LMXBs with a neutron star accretor, but
their basic methods can be applied to black hole systems as well. 
Perhaps the most powerful feature of this hot spot model is the
facility with which it can be developed and extended to more general
accretion disk geometries. For
example, our ray-tracing code could be used in conjunction with a 3D
MHD calculation of the accretion disk to simulate the time-dependent
X-ray flux and spectrum from such a disk.

\section{Ray Tracing in the Kerr Metric}
While the details of the ray tracing algorithm are presented in full in
\cite{schni03}, we give here a short summary of our methods. We
begin by dividing the image plane into regularly spaced ``pixels'' 
of equal solid angle in the observer's frame, each corresponding to a
single ray. Following the sample rays backward in time,
we calculate the original position and direction that a photon emitted
from the disk would require in order to arrive at the appropriate
position in the detector. The gravitational lensing and magnification of 
emission from the plane of the accretion disk is performed automatically
by the geodesic integration of these evenly spaced photon
trajectories, so that high magnification occurs in regions where
nearby points in the disk are projected to points with large
separation in the image plane. To model the time-varying emission from
the disk, each photon path is marked with the time delay along
the path from the observer to the emission point in the disk.

To integrate the geodesic trajectories of photons or massive
particles, we use a Hamiltonian
formalism that takes advantage of certain conserved quantities in the
dynamics: the energy, angular momentum, and mass of each particle.
By eliminating the energy $(E=-p_t)$
from the Hamiltonian, we can use the conjugate coordinate $t$ as the
integration variable, thus reducing the problem from eight phase space
dimensions to six. Another integral of the motion, Carter's constant
$Q$ \cite{carte68}, is not explicitly conserved in our equations,
but rather is used
as an independent check for the accuracy of the numerics. The resulting
equations of motion do not contain any sign 
ambiguities from turning points in the orbits, as are introduced
by many classical treatments of the geodesic equations in the Kerr
metric.

The photon trajectories are integrated backward in time from the
image plane oriented at some inclination angle $i$ with respect to the
axis of rotation for the black hole, where $i=0^\circ$ corresponds to
a face-on view of the disk and $i=90^\circ$ is an edge-on view. The
accretion disk is confined to a finite region of latitude and is
assumed to be oriented normal to the
rotation axis. The photons terminate either at the event horizon or
pass through the surfaces of colatitude $\theta={\rm const}$.
As trajectories pass through
the disk, the photon's position and momentum $(x^\mu, p_\mu)$ are
recorded for each plane intersection in order to later reconstruct an
image of the disk. The results of this paper are based primarily on
flat disks with simple hot spot perturbations, yet with the
computational methods described above, arbitrary disk geometries and
emissivity/opacity models can be simulated as well. 

\section{Hot Spot Emission}
Given the map from the accretion disk to the image plane, with each
photon bundle labeled with a distinct 4-momentum and time delay, we
can reconstruct time-varying images of the disk based on time-dependent
emission models. The simplest model we consider is a single region
of isotropic, monochromatic emission following a geodesic trajectory:
the ``hot spot'' model \cite{stell98,stell99}. We treat
the hot spot as a small region of the disk with additional emissivity
chosen to have a Gaussian distribution in local Cartesian space. We
typically take $R_{\rm spot}=0.25-0.5M$, but find the normalized light
curves and QPO power spectra to be independent of spot size and shape.

By definition the hot 
spot will have a higher temperature or density and thus greater
emissivity than the background disk, adding a small modulation to the
total flux. RXTE observations find the HFQPO X-ray modulations to have
typical amplitudes of 1-5\% of the mean flux during the outburst
\cite{remil02}. Assuming a
Shakura-Sunyaev type disk \cite{shaku73} with steady-state emissivity
$g(r) \propto r^{-2}$ and a similar scaling for the hot spot emission,
we find that small hot spots with overbrightness of $\approx 100\%$
(consistent with 3D MHD calculations \cite{hawle01}) are easily
capable of creating X-ray modulations on the order of 1\% rms.

As the disk inclination increases, the light curve goes from nearly
sinusoidal to being sharply peaked by special relativistic
beaming. Thus the shape of a QPO light curve may be used to
determine the disk inclination. Since current observational
cannot resolve the X-ray signal over individual periods as short
as 3-5 ms, instead the Fourier power spectrum can be used to
identify the harmonic features of the light curve over many
orbits. Disks with higher
inclinations will give more power in the higher harmonic frequencies,
due to the ``lighthouse'' effect, as the hot spot shoots a
high-power beam of photons toward the observer once per orbit,
approximating a periodic delta-function in time. 

\begin{figure}
\scalebox{0.45}{\includegraphics*[84,360][544,720]{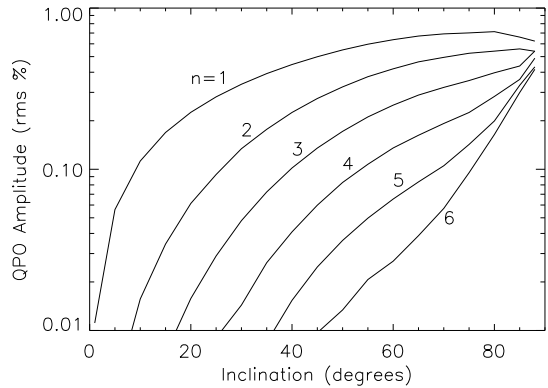}}
\caption{\label{harmonics0} Fourier amplitude $a_n({\rm rms})$ in
  higher harmonic
  frequencies $\nu_n=n\nu_\phi$ as a function of orbital inclination
  to the observer, normalized as in Eqn.\ \ref{arms}. The hot spot
  has size $R_{\rm spot}=0.5M$, an overbrightness factor of 100\%, and
  is in a circular orbit at $R_{\rm ISCO}$ around a Schwarzschild 
  black hole.}
\end{figure}

Figure \ref{harmonics0} shows the quantitative dependence of harmonic
power on disk inclination for a hot spot orbiting at the ISCO of a
Schwarzschild black hole. Predictably, as the inclination increases,
we see that both the absolute and relative amplitudes of the higher
harmonics increase, almost to the limit of a periodic delta-function
when $i\to 90^\circ$. Again, we find very little dependence of the
harmonic structure on hot spot size or shape, as long as the total
emission of the spot relative to the disk is constant. Here we have
normalized the rms amplitudes to 
the background flux from the disk with a hot spot overbrightness of
100\%. For a signal $I(t)$ with Fourier components $a_n$: 
\begin{equation}
I(t) = \sum_{n=0}^\infty a_n \cos(2\pi n t),
\end{equation}
we define the rms amplitude $a_n({\rm rms})$ in each mode $n>0$ as
\begin{equation}\label{arms}
a_n({\rm rms}) \equiv \frac{a_n}{\sqrt{2a_0}}.
\end{equation}
With this normalization, the total rms can be conveniently written 
\begin{equation}
{\rm rms} = \sqrt{\sum a_n^2({\rm rms})}.
\end{equation}

\section{Fitting QPO Data}

The high frequency QPOs observed with RXTE can be interpreted in the
context of the hot spot model by matching the model parameters to the
features of the QPO power spectrum, particularly the locations,
amplitudes, and widths of the peaks. This problem of fitting a power
spectrum to the parameters of a multi-dimensional model is analogous
to the spectacular successes that have recently been made in observing
and interpreting the perturbations in the cosmological microwave
background (CMB). Like the CMB power spectrum, the features of the QPO
power spectrum often can be isolated and fitted to one or two of the model
parameters at a time. We approach this problem in the systematic
step-by-step procedure outlined below, using the example of the Type A
QPOs from XTE J1550-564 \cite{remil02} as case study.

First, we match the frequencies of the QPO peaks to the coordinate
frequencies of specific geodesic orbits around a black hole in order
to determine the black hole mass and angular momentum. The 2:3
frequency commensurability of the two major peaks (and possibly 2:3:4
in some observations) directs our
attention to orbits with radial and azimuthal frequencies with a
ratio of $\nu_r:\nu_\phi=$1:3. The choice of the 1:3 ratio instead
of 2:3 gives the appropriate relative power in the two peaks-a
major peak at the fundamental azimuthal orbital frequency with
additional power in the sidebands at $\nu=\nu_\phi \pm
\nu_r$. Conversely, a ratio of 2:3 in the coordinate frequencies
would actually give a 1:3:5 ratio in the major power spectrum
peaks. For any value of the black hole spin $0 \le a/M \le 1$, there
exists a special radius where 
nearly circular orbits will have the appropriate 1:3 ratio in
coordinate frequencies. Then by scaling the black hole mass, we can
match the actual peak locations to the data, giving a one-dimensional
degeneracy in mass-spin parameter space. 

This degeneracy could be broken
by identifying the low frequency QPOs with the Lens-Thirring
precessional frequency at the same special radius. However, this
association is still quite speculative and has difficulty explaining
the large rms amplitude of the LFQPOs. For the present, we will
constrain our model to a region of this degeneracy limited by
the independent black hole mass determination of radial velocity
measurements, giving $8.5<M/M_\odot<11.5$ \cite{orosz02}. For the QPO
peaks at 184 Hz and 276 Hz in XTE J1550, we choose
$M=10.3M_\odot$ and $a/M=0.5$, giving a geodesic orbital radius of
$r_0 = 4.89M$, somewhat outside the ISCO at $R_{\rm ISCO}=4.23M$. 

Next, we match the total rms amplitudes in the two major peaks by
selecting an average hot spot overbrightness of 100\% and radius
$R_{\rm spot} = 0.35M$. The disk inclination is independently
determined to be $i \approx 70^\circ$ \cite{orosz02}. As explained in
the previous section, this relatively high inclination should produce
significant power in many of the higher harmonics at integer multiples
of the fundamental frequency $\nu=n\nu_\phi$. These harmonics lie well
within the sensitivity range of RXTE, yet are not detected with such
high amplitudes. Instead, we find that by shearing the hot spot into
an arc of finite length in azimuth, the power at higher frequencies is
strongly damped, due to the countering of the lighthouse effect when
the emission region is spread out over a larger portion of the orbit
\cite{schni03}. An arc length of $180^\circ$ gives the appropriate
amount of power around 184 Hz and 276 Hz while
also successfully decreasing the power at higher frequencies.

The relative power in the two major peaks is primarily determined by
the eccentricity of the geodesic orbit, as well as the arc length of
the hot spot emission region. For the Type A QPOs, even the modest
eccentricity of 0.11 is sufficient to produce significant power in the
``side-band'' at 184 Hz. Even the Type B QPOs, where the peak at 184
Hz is stronger than the fundamental peak at 276 Hz, can be explained
easily in the context of the hot spot model by shearing the arc almost
to a ring-shaped perturbation with arc length $330^\circ$. For both
cases, it is not obvious why there should be such a small amount of
predicted power at $\nu_\phi+\nu_r=$368 Hz, further advocating the
necessity of a full ray-tracing algorithm, as opposed to simple
analysis of the coordinate frequencies.

Up to this point, all of the ingredients of the hot spot model combine
to produce a power spectrum with a series of discrete peaks describing
a purely periodic light curve. In order to understand the significant
broadening of the peaks (and thus the term \textit{quasi}-periodic),
we generalize the hot spot model by replacing a single spot orbiting
indefinitely on a single trajectory by a collection of hot spots that
are continually created and destroyed with random phase along their
orbits. Each hot spot lifetime $t$ has an exponential distribution function
$f(t) \propto \exp(-t/T)$ with a probability $P=dt/T$ of being
``destroyed'' during each time step $dt$. When one hot spot is
destroyed, a new hot spot is created with the same $r_0$ but with
random orbital phase. The effect is to spread each delta-function
feature in the periodic power spectrum into a Lorentzian peak with
FWHM $\propto 1/T$. With this multiple hot spot model, the power
spectrum from XTE J1550 is closely matched by setting the typical
lifetime $T=15$ ms, or approximately four orbits in azimuth. 

\begin{figure}
\scalebox{0.45}{\includegraphics*[84,360][550,720]{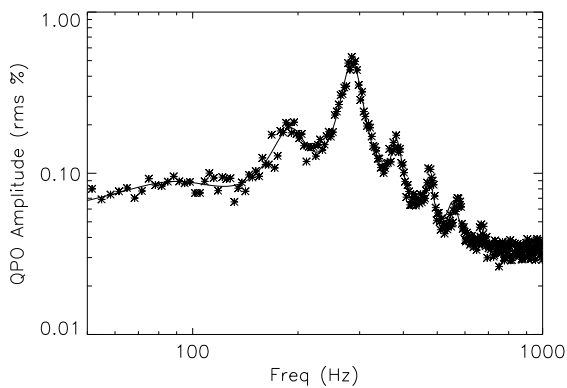}}
\caption{\label{powerspec} Simulated QPO power spectrum for a set of
model parameters described in Table 1, selected to fit the spectra of
Type A QPOs from XTE J1550-564. The solid line is a fit to the
simulated data with a series of Lorentzian peaks.}
\end{figure}

The summary of these best-fit results is shown in Table 1, and the
corresponding power spectrum (calculated from a Monte-Carlo simulation
of the light curve over many hot spot lifetimes) is shown in Figure
\ref{powerspec}. The solid curve is the best fit of a collection of
Lorentzians to the simulated data, included a small background
counting noise level with a Poisson distribution. Again, the analogy
to the CMB data is evident, and just like the with the CMB, each
successive advance in instrumentation brings us closer to a precise
determination of the model parameters. However, for the QPOs, in
addition to the power spectrum, there is also significant information
in the light curve phase, if we can achieve the necessary sensitivity
to resolve the signal in the time domain. This phase information
promises to ultimately help in distinguishing between the various
proposed explanations of the cause of the black hole QPOs.

We have also investigated the effect of various distributions in
orbital eccentricity, overbrightness, and the number of simultaneous
hot spots at any time. Interestingly, we find that the QPO power
spectrum is dependent only on the \textit{average} eccentricity and
overbrightness, and is not sensitive to the actual shape of the
distribution. Similarly, the power spectrum is independent of the
number of different hot spots in existence at any given time, sensitive
only to the average lifetime of each hot spot. 

\begin{table}
\caption{\label{best_fit} Hot spot model parameters for Type A HFQPOs
from XTE J1550-564}
\begin{tabular}{lcc}
  Parameter & & Value  \\
  \hline
  BH mass & & 10.3$M_\odot$ \\
  BH spin $a/M$ & & 0.5 \\
  disk inclination & & $70^\circ$ \\
  geodesic $r_0$ & & 4.89$M$ \\
  $R_{\rm spot}$ & & $0.35M$ \\
  eccentricity & & $0.11$ \\
  arc length & & $180^\circ$ \\
  overbrightness & & 100\% \\
  hot spot Lifetime $T$ & & 15 ms \\
  \hspace{0.5cm} (4 orbits) & & \\
  Lorentzian FWHM & & 45 Hz \\
\end{tabular}
\end{table}

One major remaining issue with the hot spot model is the preferred
location of the geodesic that gives rise to 1:3 coordinate
frequencies. Why should the orbital frequencies favor integer ratios,
and why should the preferred ratio be 1:3 and not 1:2 or 1:4? It is
possible that detailed radiation-hydrodynamic 
calculations with full general relativity will be required to answer this
question. Perhaps the non-circular orbits can only survive along
closed orbits such as these to somehow avoid destructive
intersections. For now, we are forced to leave this as
an open question unanswered by the geodesic hot spot model.

Yet with the computational framework developed above, these questions can be
answered by modeling a whole collection of hot spots and arcs
continually forming and evolving in shape and emissivity. The
particular physical parameters for these hot spots can be derived from
published MHD calculations such as \cite{hawle01}. With the basic
ray-tracing and radiation transport methods in place, it should be
possible to use our code as a ``post-processor'' to analyze the
results of these 3D simulations to simulate X-ray light curves and
spectra from a realistic accretion disk.

A final piece of the black hole QPO puzzle is the spectral behavior of
the source during outburst. \citet{remil02} have shown there is a
large range of X-ray fluxes with different relative contributions from
a power-law component and a disk bolometric component. These relative
and absolute fluxes seem to be correlated with the amount of power in
both the LFQPOs and the HFQPOs. Future work on 3-dimensional disks
and more detailed radiation transfer models should give us important
insights into understanding this spectral behavior and its relation to
the QPO power.


\begin{theacknowledgments}
We would like to thank Ron Remillard for many helpful
discussions. This work was supported by NASA grant NAG5-13306.
\end{theacknowledgments}


\bibliographystyle{aipproc}   

\bibliography{sample}

\IfFileExists{\jobname.bbl}{}
 {\typeout{}
  \typeout{******************************************}
  \typeout{** Please run "bibtex \jobname" to optain}
  \typeout{** the bibliography and then re-run LaTeX}
  \typeout{** twice to fix the references!}
  \typeout{******************************************}
  \typeout{}
 }

\end{document}